\documentclass[intlimits,twoside,a4paper]{article}
\usepackage{graphicx,psfrag}
\usepackage[T2A]{fontenc}
\usepackage[cp1251]{inputenc}
\usepackage[english]{babel}
\usepackage{soul}
\usepackage{xcolor}
\usepackage[eqsecnum]{cmpj3}

\issue{2023}{26}{1}{13507}
\doinumber{10.5488/CMP.26.13507} 

\title[Potts model with invisible states on a scale-free network]%
{Potts model with invisible states on a scale-free network}
\author[P. Sarkanych, M. Krasnytska]{ P. Sarkanych \orcid{0000-0001-6916-5004}\refaddr{label1,label2}, M. Krasnytska\orcid{0000-0002-0464-5741}\refaddr{label1,label2,label3} }

\addresses{
\addr{label1}Institute for Condensed Matter Physics of the National Academy of Sciences of Ukraine, 79011 Lviv, Ukraine 
\addr{label2}$\mathbb{L}^4$ Collaboration \& Doctoral College for the Statistical Physics of Complex Systems, 
Leipzig-Lorraine-Lviv-Coventry, Europe
\addr{label3} Laboratoire de Physique et Chimie Th\'eoriques,  Universit\'e de Lorraine, Vandoeuvre-les-Nancy, 54506, France}


\date{Received October 31, 2022, in final form November 21, 2022}

\begin{document}
\maketitle
\begin{abstract}
Different models are proposed to understand magnetic phase transitions  through the prism of competition between the energy and the entropy.  One of such models is a $q$-state Potts model with invisible states. This model introduces $r$ invisible states such that if a spin lies in one of them, it does not interact with the rest states.  
We consider such a model using the mean field approximation  on an annealed scale-free network where the probability of a randomly chosen vertex having a degree $k$ is governed by the power-law $P(k)\propto k^{-\lambda}$.  
Our results confirm that $q$, $r$ and $\lambda$ play a role of global parameters that influence the critical behaviour of the system. Depending on their values, the phase diagram is divided into three regions with different critical behaviours. 
However, the topological influence, presented by the marginal value of $\lambda_c(q)$, has proven to be dominant over the entropic influence, governed by the number of invisible states $r$.
 
\printkeywords
%
\end{abstract}

{\em Our first paper on Potts model with invisible states was published together with Bertrand Berche. This paper is submitted for a special issue: ``Complexity and collective behaviour: Solids, Fields, and Data'' on the occasion of Bertrand Berche's 60th birthday. By the present work we would like to thank the jubilee and wish him an inspiration and interesting research topics to study.}

\section{Introduction}
\label{I}

Critical properties of a system are defined by the topology of the space this system is placed into. The nearest neighbours ferromagnetic Ising model can be considered as the textbook example of this. When it is placed on a one dimensional chain, it exhibits no phase transition \cite{Ising}, while in two and three dimensions it undergoes a second order phase transition with critical exponents dependent on the dimensionality~\cite{Onsager,El-Showk}. 
An interesting behaviour is observed if this model is placed on a network. 
The phase transition studies on networks are motivated mainly by the variety of real systems and objects described by the topology of a network or a graph: 
from sociophysics and biophysics to transport networks and their resistance to attacks  \cite{networks_1,networks_2,networks_3,networks_4,networks_5,Dorogovtsev08}.  
For example, individuals of social networks may be considered as nodes of the graph and individuals' opinions as a value of the spin variable for a given node.
The  dimensionality is not defined for networks in a traditional Euclidean sense, although there are some other characteristics that describe the topology, one of them being a node degree distribution. The most studied is the class of scale-free networks, where the probability of a particular node having $k$ neighbours is governed by a power-law decaying node degree distribution:
\begin{equation}
\label{pofk}
P(k)\sim 1 /{k^\lambda}, \quad k \to \infty  .
\end{equation} 
The decay exponent $\lambda$ in this case determines the properties of a phase transition in a similar way the dimensionality for regular latices \cite{Leone02,Igloi02,Dorogovtsev02}. Therefore, tuning $\lambda$ allows us to analyse how the properties of a phase transition depend on the topology of a system. 

Besides the topology, there are other factors that play a significant role in the properties of a phase transition. One of those is entropy. Phase transitions, in general, are often explained through the lens of the energy-entropy interplay. Shifting this equilibrium to one of the sides can drastically change the properties of the phase transition. 
One of the models, that allows us to investigate the energy-entropy interplay is the Potts model with invisible states \cite{Tamura10,Tamura11,Tanaka11a}. 
It was proposed to explain why the phase transition in a system with $q$-fold symmetry undergoes a different scenario than predicted theoretically~\cite{Tamura10}. 
It differs from the original $q$ state Potts model by adding $r$ invisible states, in which a spin does not interact with the rest of the system. 
Obviously, such generalisation does not affect the interaction energy spectrum of a system but changes the number of configurations by which each energy level is occupied. 
In turn, changing the density of states leads to a change in entropy.  
Adding invisible states can make the phase transition ``sharper'', or even can change the second order phase transition into the first order transition. 
For example, for the Ising model with invisible states on a complete graph, the marginal value $r_c=3.62$ separates the first and second order regimes \cite{Tamura10,Tamura11,Tanaka11a}. A series of analytical studies for different lattice systems were provided \cite{Johnston13,Mori12,Enter11a,Enter11b,Ananikian13}.

Some exact results for the model can be obtained for 1D systems.  
A positive number of invisible states does not lead to the phase transition, while a negative number of invisible states induces a phase transition at positive temperature \cite{Sarkanych17}. 
A conjecture on the relationship between critical residual entropy and finite temperature pseudo-transitions of one-dimensional spin models was analysed \cite{Rojas20}. 
An interesting fact is that such models break the Perron--Frobenius theorem~\cite{Rojas20,Ninio76,Sarkanych17}: free energy becomes non-analytical at the phase transition temperature, or equivalently some elements of transfer-matrix become null (which corresponds
to an infinite energy).

The spin models with `invisible states' became a subject of recent research and different applications.   
The Potts model with invisible states is used in elementary coordination-type games by extending equivalent models with neutral strategies \cite{Kiraly19}. 
An Ising model on a complete graph  
with the so-called ``molecule states'' contributing to the entropy, rather than the  interaction energy, was studied by means of MFA and Monte Carlo. 
This model was proposed to study the structure-forming systems ranging from the atoms that build molecules to the self-assembly of colloidal amphibolic particles. 
It was shown that the model exhibits a first-order phase transition when the number of ``molecule states'' is high \cite{Korbel21}. 
An interesting is the case of a hybrid Potts model \cite{Schreiber22}. Within $q_c$ states, it introduces two subsets of states, that can be occupied with different probability.
Namely, each spin with a probability $p$ lies in one of the $q_0 \leqslant q_c$ `strong' states and with probability $(1-p)$ in the rest $q_c-q_0$ `weak' states. It was shown that for a given $q$, there exists a concentration $p$ such that, when cooled, the system first exhibits a continuous
transition and then a discontinuous transition at a lower temperature. 

In the previous research we considered the critical behaviour for the $q$-state Potts model with invisible states on a complete graph  \cite{Krasnytska16} being interested in the behaviour in the second order phase transition region $1\leqslant q \leqslant 2$. 
We observed and calculated two marginal values of $r$ separating regions with different criticality:  for small values of $r<r_{c1}$ the system undergoes only the second order phase transition. 
For large $r>r_{c2}$, there is only the first order phase transition, while in-between these two regions we observed the coexistence of both phase transitions at different temperatures. 
We have also seen that in the limiting case $q=2$, these two marginal values coincide $r_{c1}=r_{c2}$. 
Then, we analyzed the $q=2$-state Potts (Ising) model with invisible states on a scale-free network \cite{Sarkanych19}. 
Similarly to the complete graph case, there are two marginal dimensions. 
However, on the scale-free networks, marginal dimensions are $\lambda$-dependent.  
Therefore, it is interesting to see how this effect manifests itself for all values of $q$. 
In order to investigate the effect that the two above-mentioned factors have on a phase transition if included simultaneously, in this paper we consider the Potts model with invisible states on an annealed scale-free network.   
For the annealed random networks, the edges are in thermodynamic equilibrium, and thus can fluctuate in time. This is not the case for quenched networks, where the edges are considered static. In this paper we consider only the annealed networks.

The rest of the paper is organised as follows: in section \ref{II} we shortly describe the model and the method used for the study;  in section \ref{III} we present the main results obtained for $1<q\leqslant2$ and $q>2$; conclusions are given in section \ref{IV}.

\newpage

\section{Model and method}
\label{II}
For the Potts model with invisible states, the interaction between two neighbouring spins contributes to the energy only if both of the spins are in the same `visible' state. The state of each of the spins is described by a spin variable $S_i$ that can take on $(q+r)$ different values $S_i=(1,\dots,q,q+ 1,\dots,q+r)$. Hence, the first $q$ states are considered to be `visible', while the rest $r$ are invisible. 
Introducing the external magnetic field $h$ to favour the first visible state $S_i=1$, a  ferromagnetic $q+r$-state Potts model Hamiltonian on an arbitrary graph  reads \cite{Krasnytska16}:
\begin{equation}\label{1a}
- H(q,r)=\sum_{<i,j>}J _{ij}\sum_{\alpha=1}^q \delta
_{S_i,\alpha}\delta _{\alpha,S_j}+ h\, \sum _{i=1}^N  \delta_{S_i,1}. 
\end{equation}
The first summation in (\ref{1a}) is performed over all pairs of spins in the network of $N$ nodes, while the second sum in the first term requires both of the interacting spins to be in the same visible state. Therefore, only visible states contribute to the interactions between the spins.
Two spins are only interacting if they are connected by the link of the graph our model is considered on. Therefore, we can consider coupling~$J_{ij}$ to be proportional to the adjacency matrix elements of the graph $J_{ij}=J A_{ij}$. The elements of the latter are equal to $1$ when there is a link between nodes $i$ and $j$ and $0$ otherwise.  
Within the mean-field approach, we also assume that the
network on which our model is considered, is a completely random annealed network~\cite{Lee09}. 
In this case, the adjacency matrix is equal to the probability of two nodes being connected~$p_{ij}$. 
For an annealed network, this probability is linearly proportional to their degrees (the number of nearest neighbours) \cite{Lee09}; thus, we obtain: 
\begin{equation}\label{11a}
J_{ij} =Jp_{ij}=J \cfrac{k_i k_j}{N\bar{k}},
\end{equation} 
where $k_i$, $\bar{k}$ are the degree of the node $i$ and an average node degree, respectively.
 
We use a variant of a mean-field approach (MFA)\footnote{The exactness of the mean-field results for the Potts model with invisible states was analysed in reference~\cite{Mori12}.} with local variables as suggested for a standard Potts model on a  network \cite{Igloi02,Krasnytska13}. 
The main idea is to rewrite the interaction term in the Hamiltonian~(\ref{1a}) in a way that does not contain the product of two spin variables. 
To this end, we represent each Kronecker-delta term as a sum of its mean value and deviation from that mean. 
We introduce three local thermodynamic averages $\mu_i,\nu_{1i},\nu_{2i}$ to distinguish between the state favoured by the magnetic field, other visible states and invisible states:
\begin{align}
	\begin{split}\label{2'}
	 \langle \delta_{S_i,\alpha} \rangle= & 
		\begin{cases}
                \mu_i,  & \alpha=1, \\
                \nu_{1i}, & \alpha=2, \dots, q, \\
                \nu_{2i},  & \alpha=q+1, \dots, r.
  	  \end{cases}
	\end{split}
\end{align}
Here, the averaging is performed with respect to the Hamiltonian (\ref{1a})
\begin{equation}\label{3aa}
\langle\dots\rangle = \frac{1}{\cal Z}{\rm Tr}\, (\dots) \,
\re^{-\beta{\cal H}}, \quad  {\mbox{with}} \quad 
 {\cal Z} =  {\rm Tr}\, \re^{-\beta{\cal H}},
\end{equation}
in the thermodynamic limit, where $\beta$ is the  inverse temperature and the trace is taken over all
possible spin configurations (taking into account invisible states as well). 
The introduced averages have predictable low-temperature and high-temperature asymptotics shown in table \ref{tab1}.

\begin{table}[h]
	\caption{Low-temperature and high-temperature asymptotics of the thermodynamic
		averages, equation~(\ref{2'}), and for the order parameters, equation~\eqref{4'}.
		\label{tab1}}
	\vspace{0.2ex}
	\begin{center}
		\begin{tabular}{|c|c|c|c|}
			\hline
			$\beta \rightarrow \infty$ & $\mu_i=1$ & $\nu_{1i}=0$ & $\nu_{2i}=0$  \\  \hline
			\hline
			$\beta\rightarrow 0$ & $\mu_i={1}/({q+r})$ & $\nu_{1i}={1}/({q+r})$ & $\nu_{2i}={1}/({q+r})$  \\
			\hline
		\end{tabular}
	\end{center}
\end{table}

Assuming that the deviations from the averages are small, we neglect quadratic deviation terms  to obtain the mean-field Hamiltonian:
\begin{equation}\label{9''}
- H(q,r)=\sum_{<i,j>}J_{ij}\Bigg[\mu_i\big(2\delta
_{1,S_j}-\mu_j\big)+   \sum_{\alpha=2}^q \big(2\delta
_{\alpha,S_i}-\nu_{1i}\big)\nu_{1j}\Bigg]+h\sum_i \delta_{S_i,1}.
\end{equation} 
Taking into account the normalization condition for the averages:
\begin{equation}\label{4}
\mu_i+(q-1)\nu_1+r\nu_2=1,
\end{equation} 
we define two local order parameters that satisfy low-temperature and high-temperature asymptotics: they vanish for $\beta \rightarrow 0$ and are equal to one for $\beta \rightarrow \infty$:
\begin{equation}\label{4'}
m_{1i}=\mu_i-\nu_{1i}, \quad
m_{2i}=\mu_i-\nu_{2i}.
\end{equation}
The  Potts model with invisible states on a  
network is described by two global order parameters: $m_1$ and~$m_2$.  We introduce them as a linear combination of the local order parameters defined on each of the nodes 
with weights proportional to the degree of a node (see \cite{Sarkanych19} for more details):
\begin{equation}\label{11b}
m_1=\frac{\sum_i k_i m_{1i}}{\sum_i k_i}, \qquad
m_2=\frac{\sum_i k_i m_{2i}}{\sum_i k_i}.
\end{equation}
Considering the free energy per site  in the thermodynamic limit ($N\to\infty$) we
obtain~\cite{Sarkanych19}:
\begin{eqnarray} \nonumber
    f(m_1,m_2)&=&\frac{\bar{k}}{(q+r)^2}\Big\{\left[rm_2+1+(q-1)m_1\right]^2+ (q-1)[rm_2+1  -(r+1)m_1]^2\Big\}    \\ && \nonumber
   -\frac{1}{\beta}\int_2^\infty \rd k P(k)
\ln\Bigg(\exp\Bigg\{\beta\Bigg[h+\frac{k}{q+r}(m_1(q-1)+1+rm_2)\Bigg]\Bigg\}
  \\ &&  \label{ff0}  +(q-1)\exp\Bigg\{\frac{\beta
  k}{q+r}[m_2r+1-(r+1)m_1]\Bigg\}+r\Bigg),  
\end{eqnarray}
where $P(k)$ is a node degree distribution (\ref{pofk}); $k$, $\bar{k}$ are the node degree and an average node degree, respectively\footnote{In order  to have a giant connected component for a given scale-free network, the lower integration boundary should be set to $k_{\rm {min}}=2$ \cite{Aiello2000}.}. 
The resulting expression for the free energy depends on two global order parameters $m_1$ and $m_2$, that describe the state of the system and model parameters, such as the numbers of states $q$ and~$r$, temperature $\beta$ and the decay exponent $\lambda$ (which determines the topology of a system).   
 

\section{Results}
\label{III}

\begin{figure}[h!]
	\center
    \includegraphics[width=0.55\columnwidth]{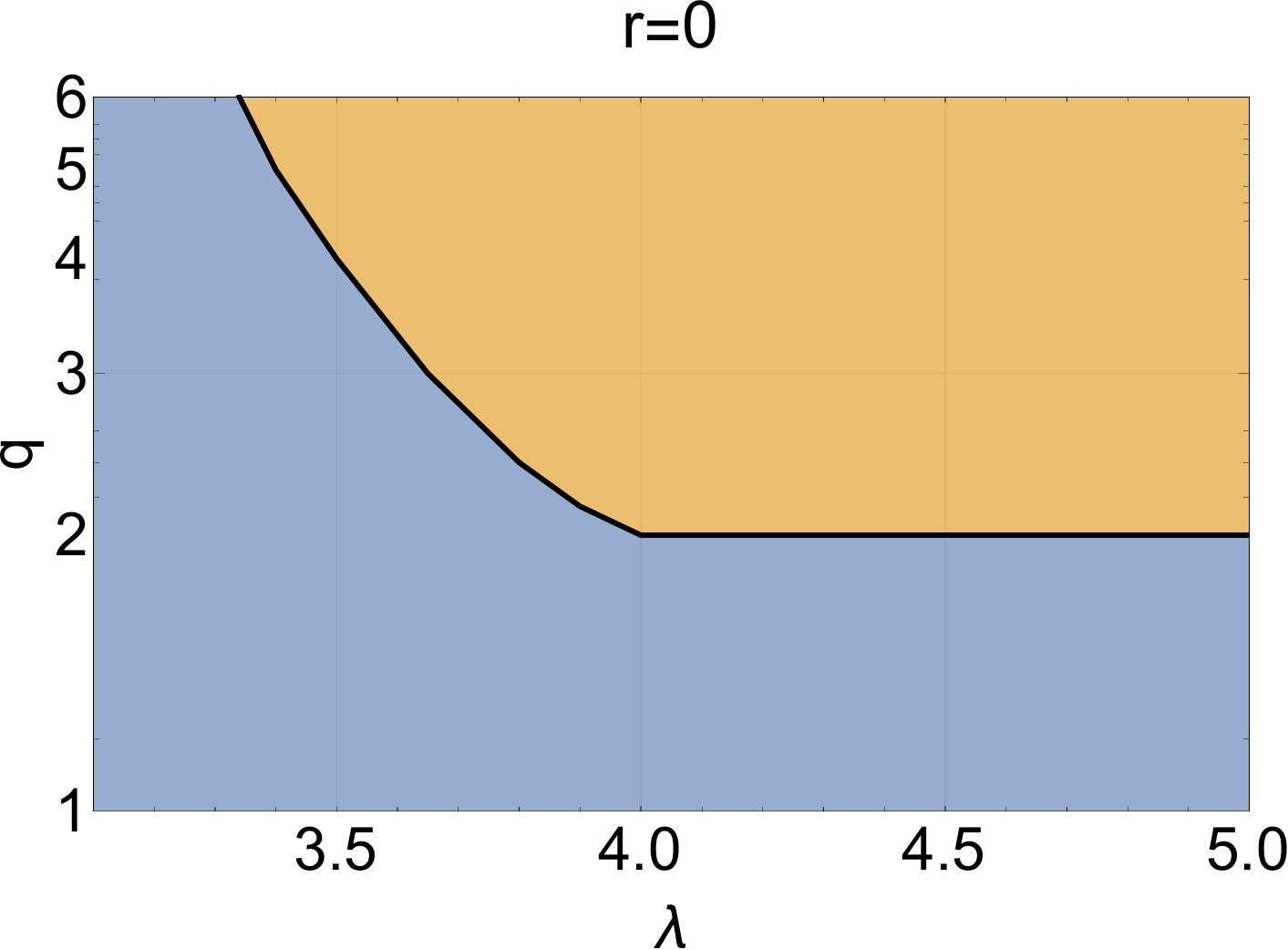}
    \caption{\label{fig6}(Colour online) Phase diagram of the standard Potts model ($r=0$) on a scale-free network. The system obeys either the first or the second order phase transition. For $1<q\leqslant 2$, the second order phase transition is observed, while for $q>2$, the critical value of $\lambda_c$ should be taken into account. For each~$q$  at $\lambda>\lambda_c (q)$,  the first order phase transition occurs, while for a smaller $\lambda \leqslant \lambda_c (q)$, the second order phase transition occurs.  }\label{fig1}
\end{figure}

In the limit $r\to 0$, our model reduces to the standard
Potts model on a scale-free network, which was already analysed within the MFA approach \cite{Leone02,Igloi02,Dorogovtsev02,Krasnytska13}. In this case, the phase diagram in the $(q,\lambda)$-plane is characterised by different criticality in different regions. For small $\lambda$ ($2<\lambda\leqslant3$), the network is strongly correlated and remains ordered at any finite temperature  due to the fact that there are too many nodes with high degrees, making the network too correlated for the occurance of the phase transition to a paramagnetic state.  
For $\lambda>3$, we can observe  different scenarios depending on the value of $q$. The region $1\leqslant q\leqslant 2$ is characterized by the second order phase transition behaviour, although with three different sets of critical exponents corresponding to the regions  $q=1$, $1<q<2$, $q=2$.
The remaining case $q>2$ is the most interesting, since either the second or the first order phase transition occurs depending on the value of $\lambda$ \cite{Leone02,Igloi02,Dorogovtsev02,Krasnytska13}. For an illustration see figure~\ref{fig1} where the phase diagram of the standard Potts model is shown. 
   
In contrast to the Ising model on a scale-free network, the Potts model obeys a richer phase diagram and  regulates the order of the phase transition.  
Thus, there is a marginal dimension $\lambda_c (q)$ separating the regions with different criticality. 
Our goal here is to investigate the role of the interplay of the parameter~$\lambda$ and the number of invisible states $r$ on the critical behaviour of the Potts model on a scale-free  network in different regions of $q$, $r$ and $\lambda$. Since for the ordinary Potts model ($r=0$), critical behaviour depends on $q$, we expect the same situation when the invisible states are added.
Within our analysis, we were unable to consider the limit $q\to 1$.
Therefore, we focus on a spontaneous ($h=0$) magnetization in three regions: $1<q<2$, $q=2$ and $q>2$.

This is a nontrivial task to analyze the critical behaviour of the model using analytical methods because the free energy of the system depends on a huge number of different parameters.
To investigate the critical behaviour, we adopt the simulated annealing minimization method \cite{SimulatedAnnealing}\footnote{Compared to our previous analysis in \cite{Sarkanych19} we improved the algorithm for a better search for the global minima on the boundary of the search region. This led to improvements in finding the minima at low values of $m_1$ and consequently to better estimates for $r_{c1}$ and $r_{c2}$.} to numerically obtain the values of the order parameters $m_1$ and $m_2$, that lead to the lowest free energy. 
To this end, we fix the numbers of visible $q$ and invisible $r$ states, parameter $\lambda$ and minimize the expression for the free energy with respect to two global order parameters [$m_1$, $m_2$ defined in (\ref{11b})] while changing the temperature. 
The obtained temperature dependencies of the order parameters allow us to distinguish between the first and the second order phase transition regimes. 
However, only the order parameter $m_1$ has a physical interpretation as a quantity that appears below the transition point and breaks the system symmetry. The second order parameter $m_2$ disappears only if there are no invisible states $r=0$ \cite{Sarkanych19,Krasnytska16}. If $m_1$ smoothly vanishes as the temperature increases ($\beta$ decreases), then the transition is of the second order. On the contrary, if there are discontinuities in $m_1(\beta)$, then the system undergoes the first order phase transition.

\subsection{Strongly correlated networks: $2<\lambda\leqslant3$} \label{III.1}

\begin{figure}[h!]
	\center
    \includegraphics[width=0.45\columnwidth]{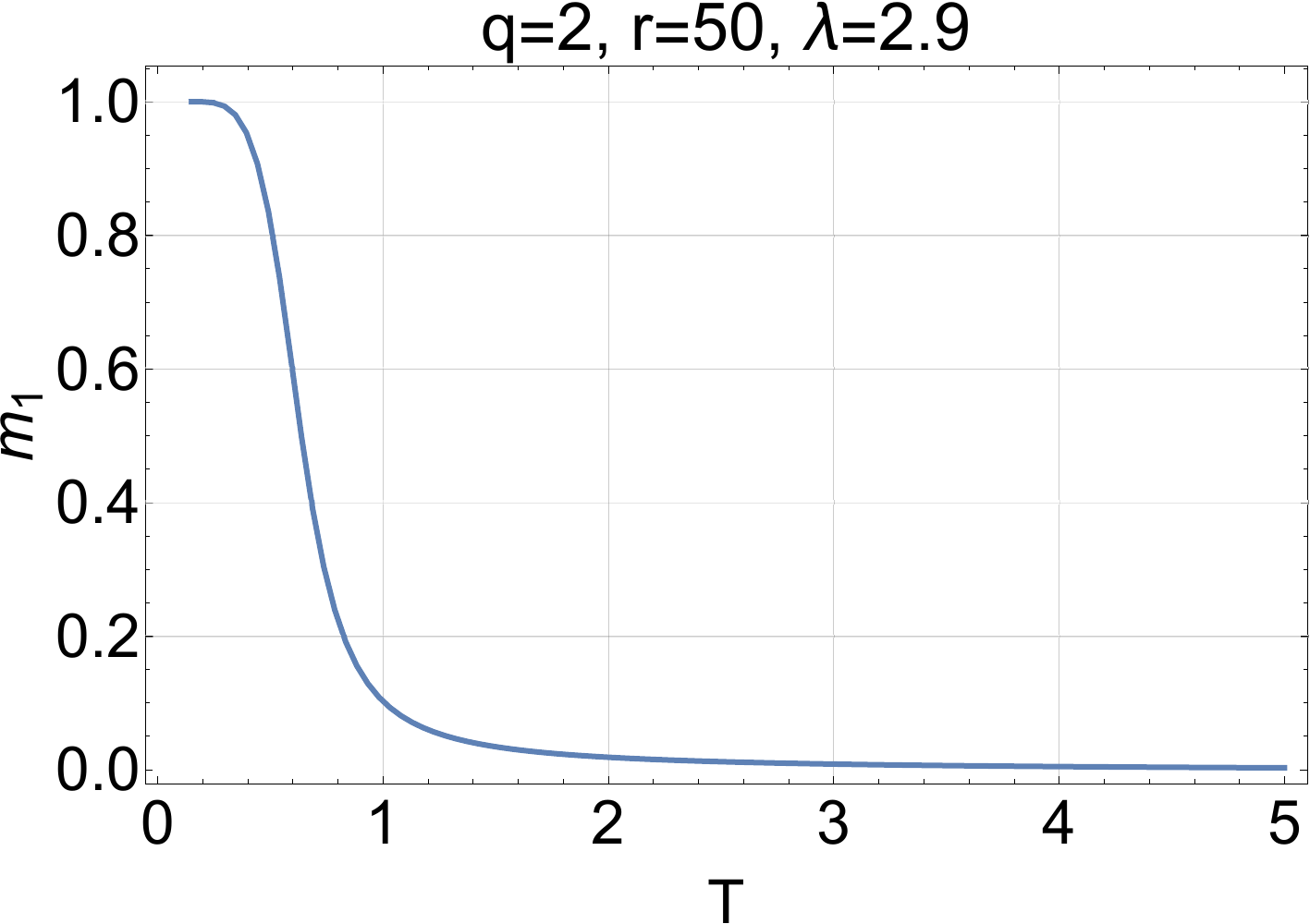} \hspace{0.5cm}
    \includegraphics[width=0.45\columnwidth]{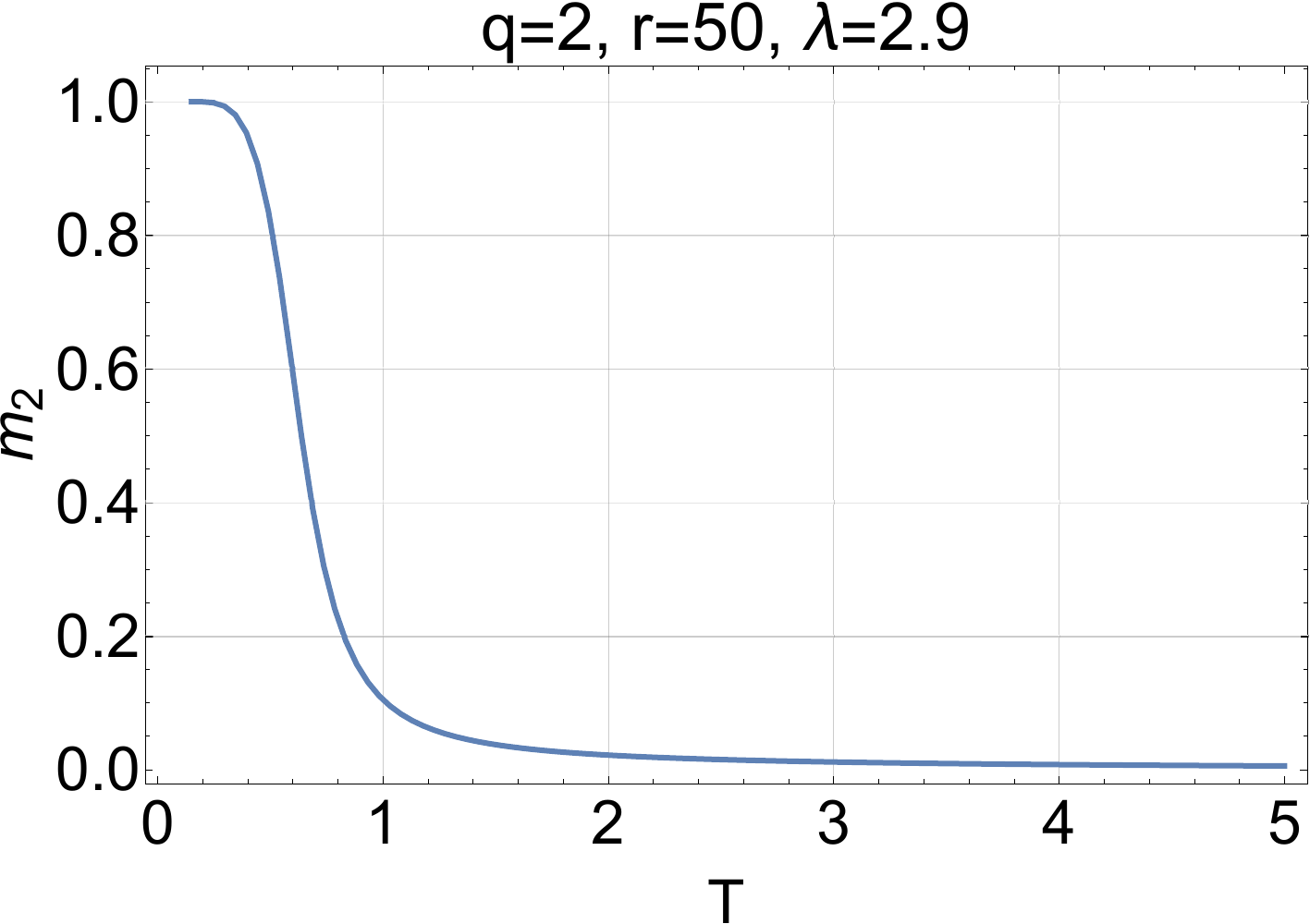}
    \caption{\label{lowl}(Colour online) Order parameter dependency on temperature for the Potts model with $q=2$ visible and $r=50$ invisible states on a scale-free network with degree distribution exponent $\lambda=2.9$. Both order parameters decay with temperature but never vanish. This signals the absence of the phase transition.}
\end{figure}

We start our analysis with strongly correlated networks, i.e., where $\lambda\leqslant3$. 
For the ordinary Potts model on a scale-free network, it was shown that in this case the system remains ordered at any temperature, meaning that there is no phase transition. 
This happens because the degree distribution vanishes slowly and many nodes in the network have high degrees forming a strongly connected and clustered structure. 
This type of behaviour is also observed in the Potts model with invisible states: no large amount of invisible states is capable of breaking the order.
To illustrate this, in figure~\ref{lowl} we present the dependency of both order parameters on the temperature at large $r$.
These smooth curves indicate that there is no phase transition for the network with a slow degree distribution decay, similar to what is observed for the standard Potts model.

\subsection{Region $1<q \leqslant 2$}\label{III.2}

\begin{figure}[h!]   
	\center
	\includegraphics[width=0.45\columnwidth]{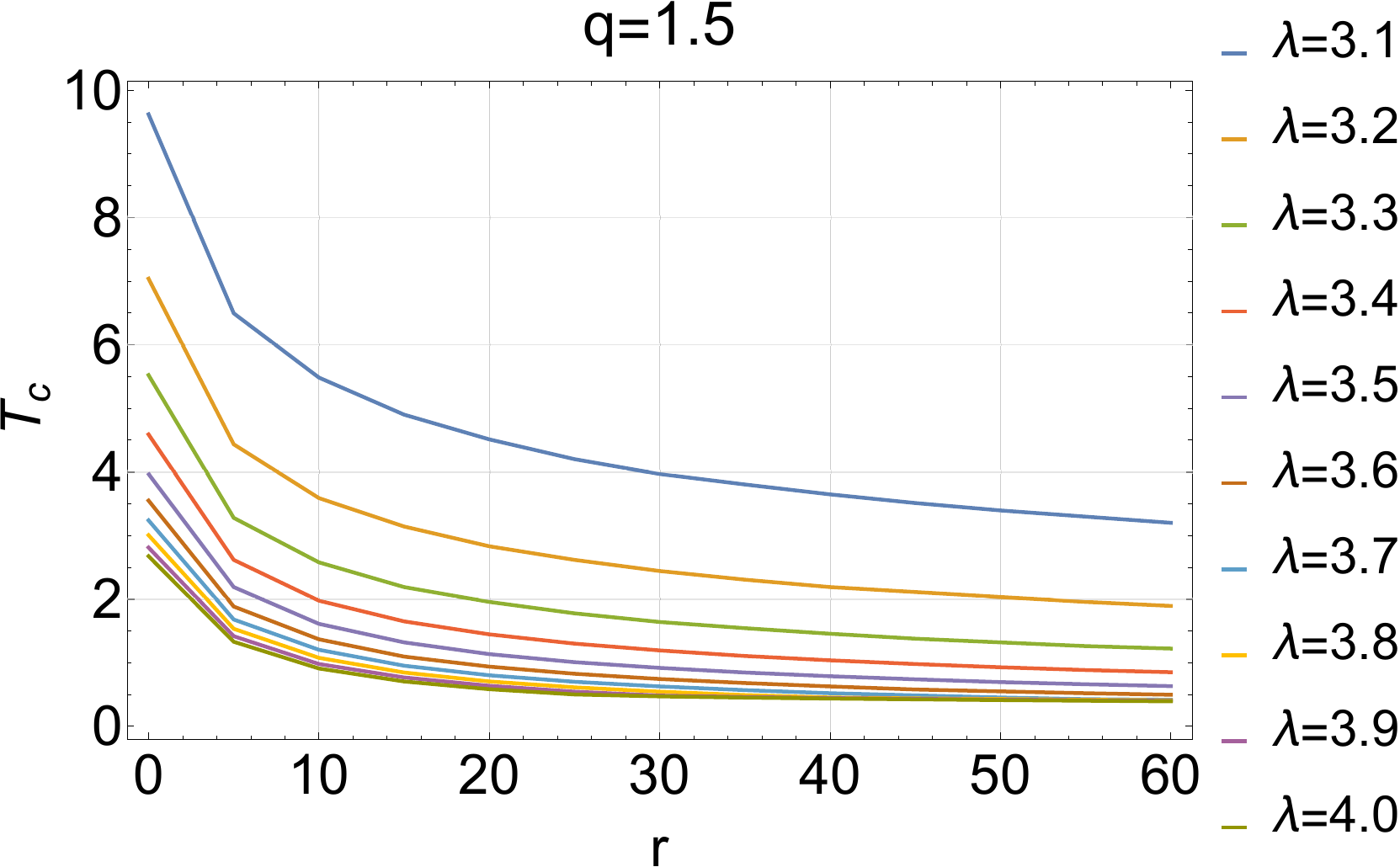} \hspace{0.5cm}
    \includegraphics[width=0.45\columnwidth]{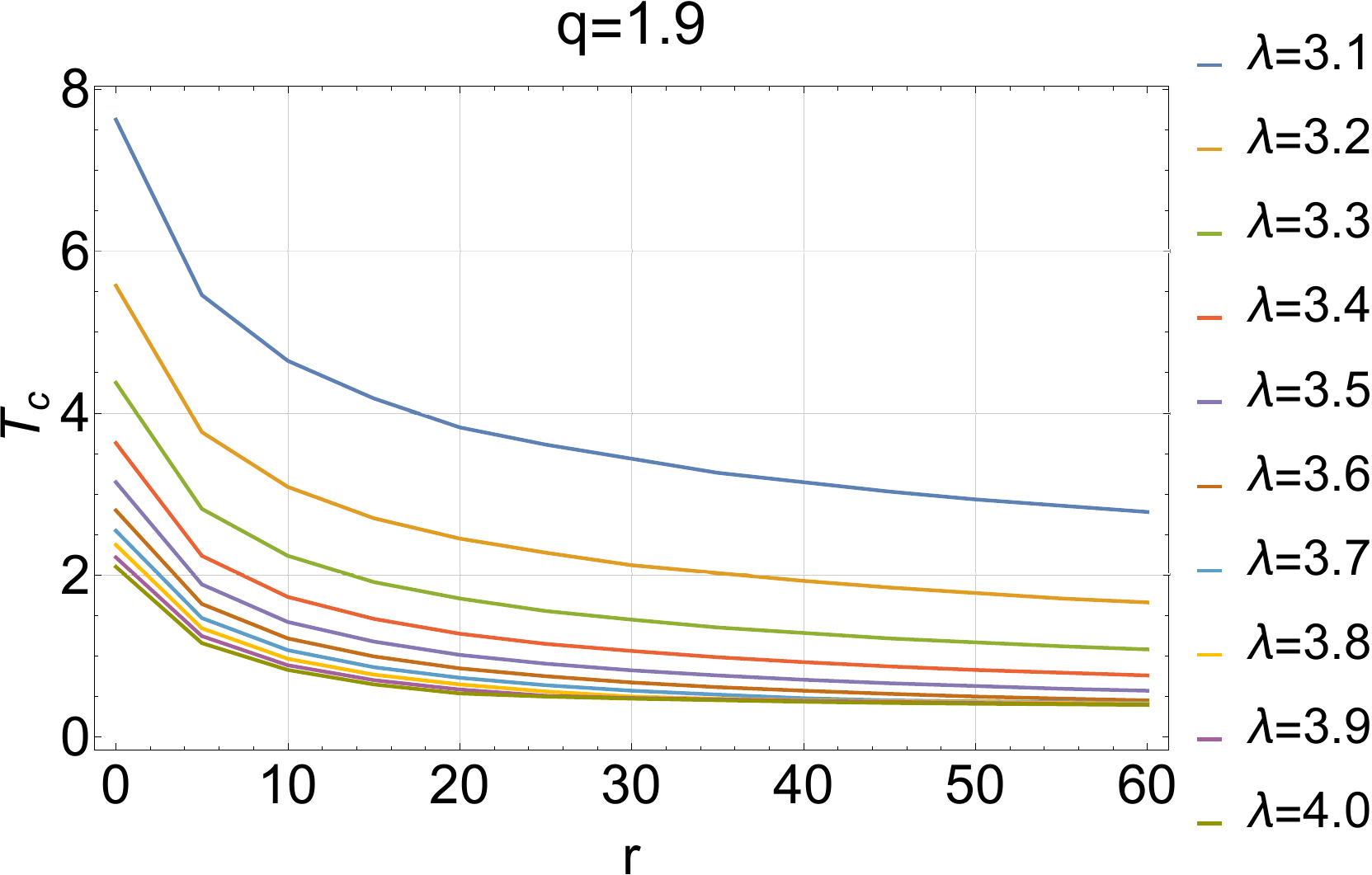}
    \caption{\label{tclowq}(Colour online) Critical temperature as a function of the number of invisible states $r$ for different values of $\lambda$. Similarly to the case $q=2$, critical temperature decreases with the increase in both $r$ and $\lambda$. In addition, an increase in $q$ lowers the critical temperature. We present only values $3<\lambda\leqslant 4$ to make the plot less cluttered, although the trends are the same in the region $\lambda>4$.}
\end{figure}

In this subsection, we present the results for the $q=1.5$ and $q=1.9$ Potts model to illustrate how the addition of invisible states affects the critical behaviour.  
This case in many aspects is similar to the Ising case ($q=2$) considered in \cite{Sarkanych19}. 
We start by examining the critical temperature. In figure~\ref{tclowq} we present the values of critical temperature for different values of $q$, $r$ and $\lambda$.
We see that the addition of invisible states as well as an increase in $\lambda$ decreases the critical temperature. 
The reason for the former effect is that the number of invisible states corresponds to the entropy. Therefore, with more entropy it is easier to break the ordering.
The increase in $\lambda$ translates into a lower number of hubs that play a significant role in the network connectivity and, therefore, in the ordering. We have already observed this in the previous subsection.

As we have already mentioned, the  order parameter $m_2$ stays finite for all finite temperatures and at $r=0$, $m_2=0$, and we only have a single order parameter $m_1$. Therefore, to detect the phase transition we will only track the order parameter $m_1$. However, both $m_1$ and $m_2$ can be used to distinguish between the first and the second-order regimes as the plots demonstrate.

\begin{figure}[h!]
	\center
    \includegraphics[width=0.45\columnwidth]{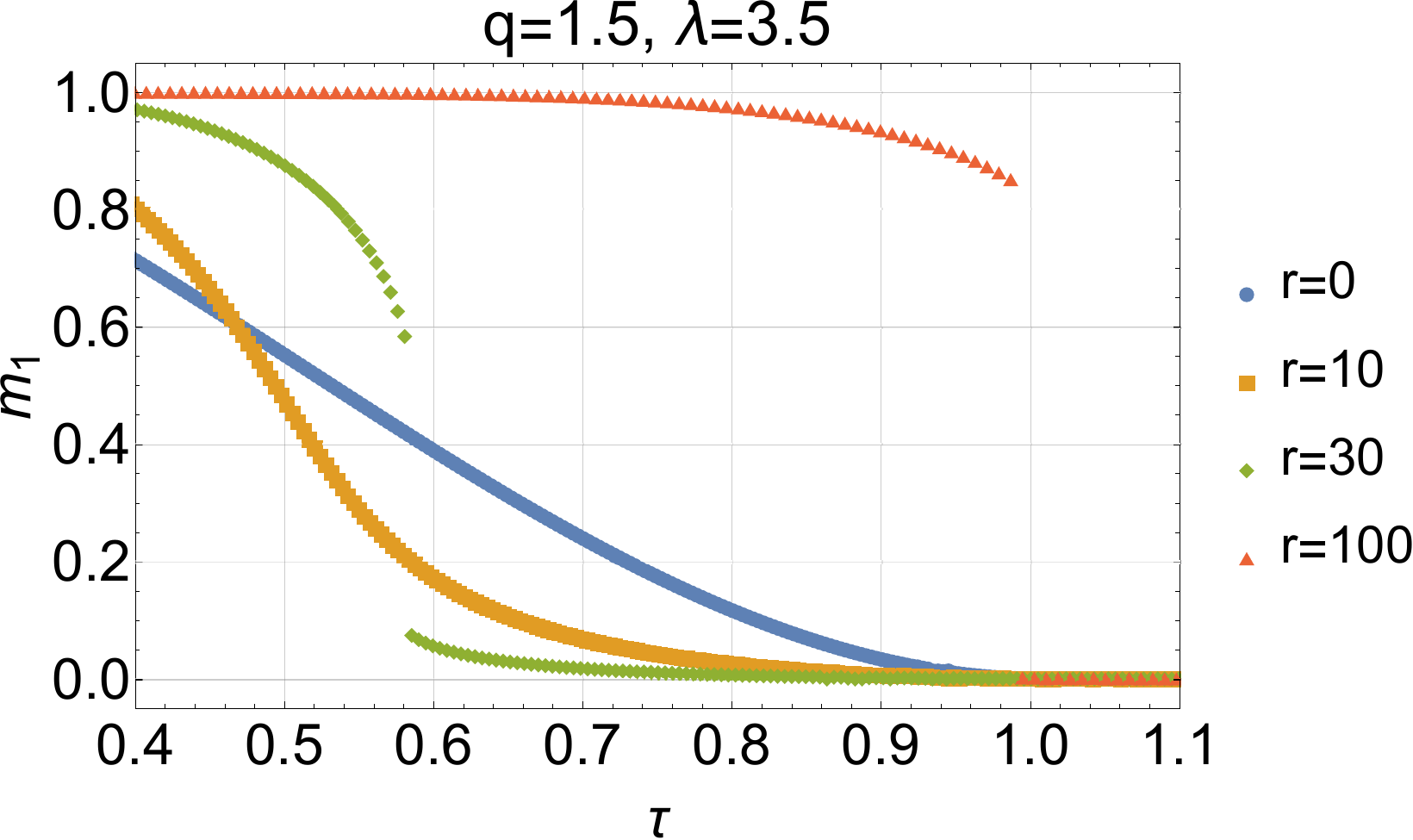} \hspace{0.5cm}
    \includegraphics[width=0.45\columnwidth]{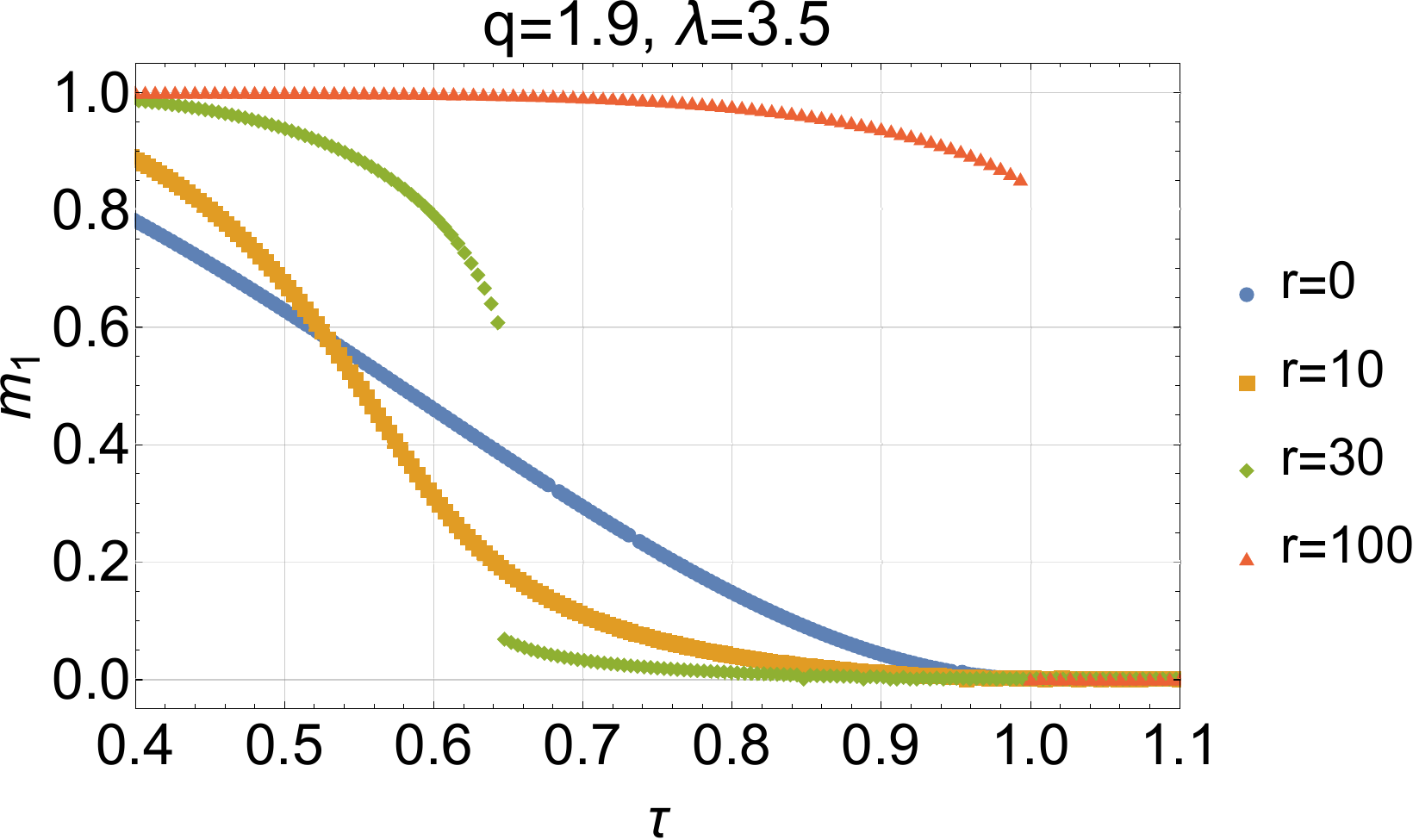}
    \caption{\label{m1lowq}(Colour online) The order parameter $m_1$ dependency on a reduced temperature $\tau=T/T_c$. The left-hand plot shows data for $q=1.5$ while the right-hand one is for $q=1.9$. 
    For low values of $r$, the system undergoes the second order phase transition, while for large $r$ there is only the first order transition. Models with an intermediate number of invisible states are characterized by two phase transitions: the first order phase transition between two partially ordered states at a lower temperature and the second order transition at a higher temperature.}
\end{figure}  

\begin{figure}[h!]
 \center
    \includegraphics[width=0.45\columnwidth]{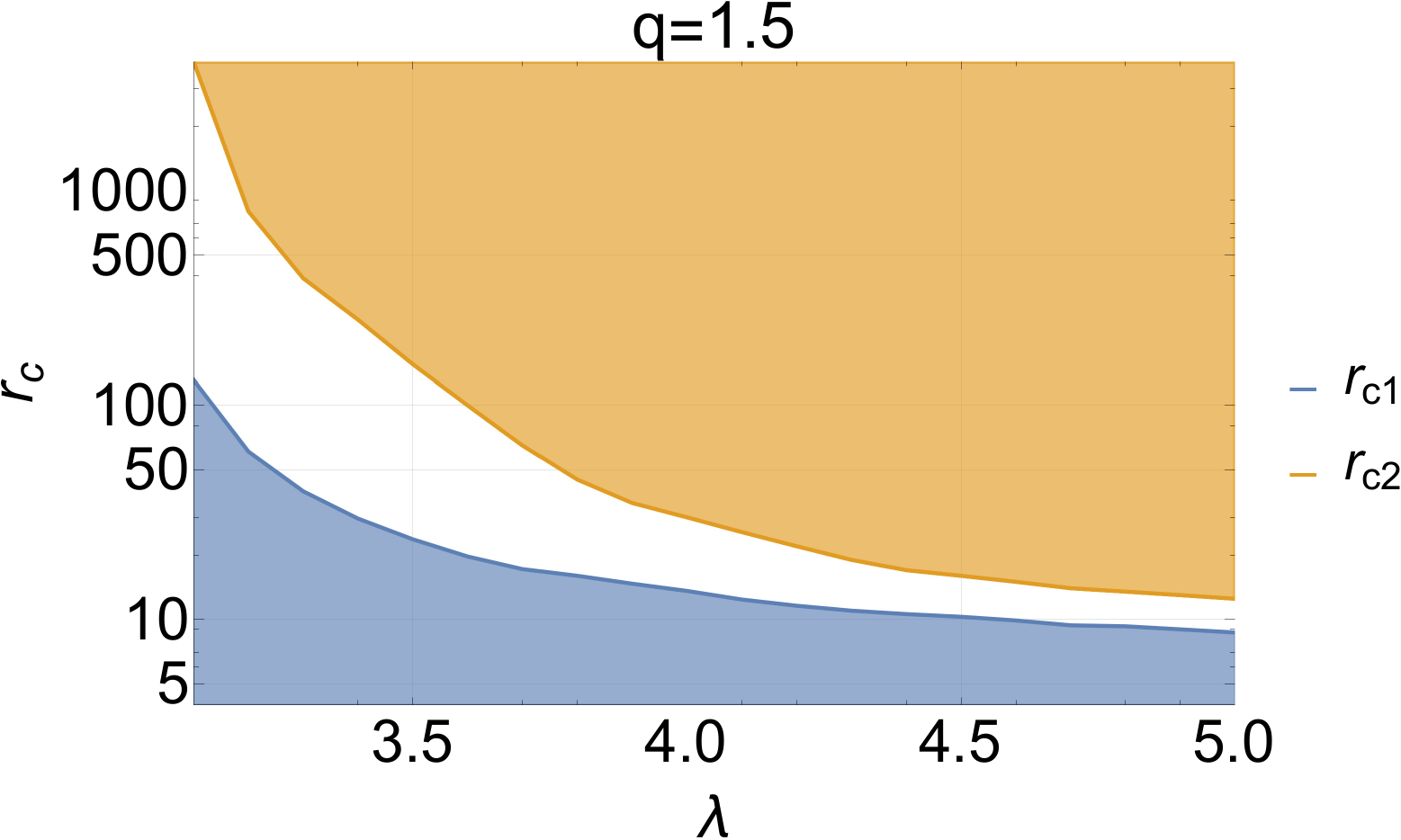} \hspace{0.5cm} \includegraphics[width=0.45\columnwidth]{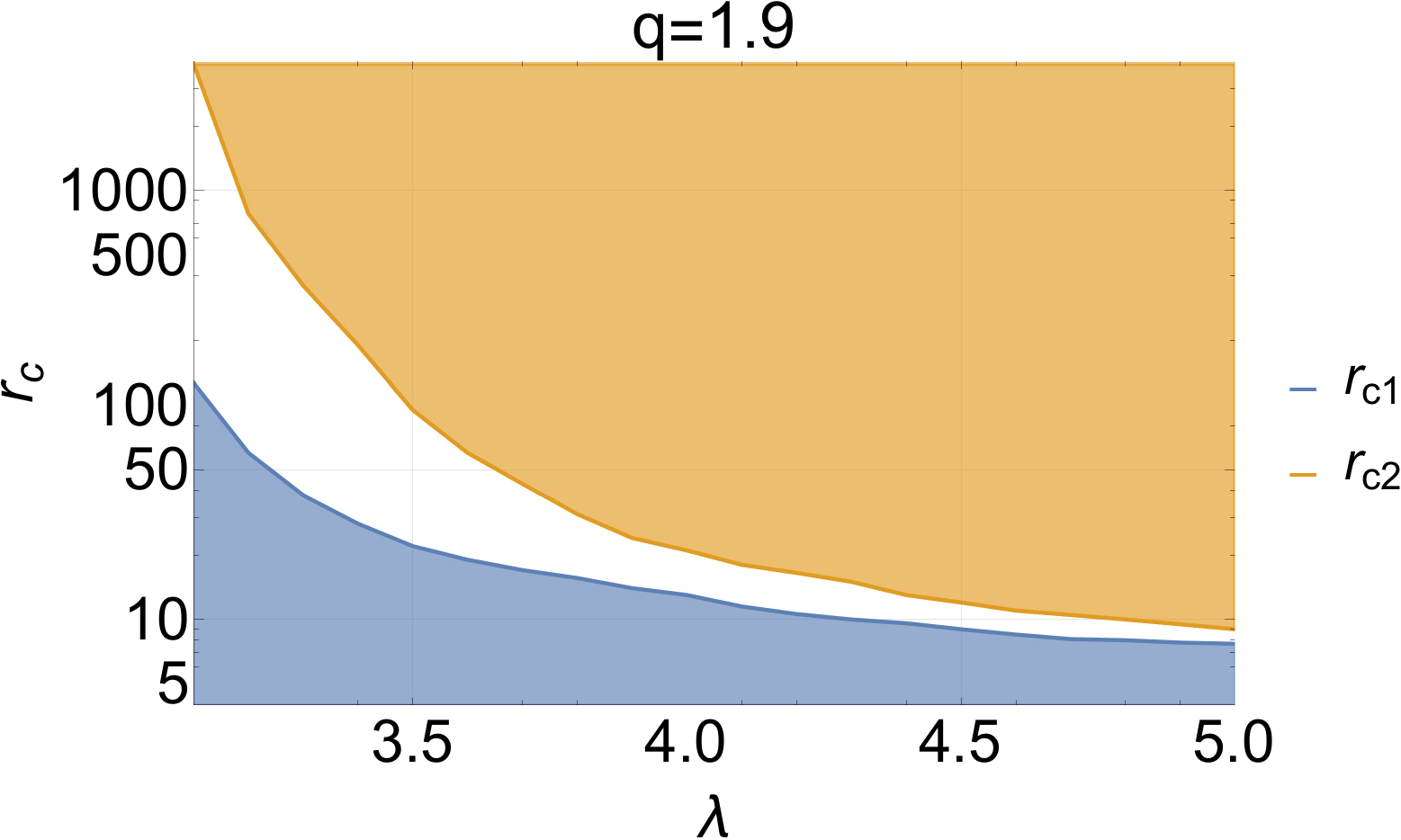}
    \caption{\label{fig3}(Colour online) Phase diagram of the Potts model with invisible states on a scale free network for $q=1.5$ (the left-hand plot) and  $q=1.9$ (the right-hand plot) at different $r$ and $\lambda>3$.  The three regions, presented here, differ in critical behaviour. In the lower (blue) region the system undergoes only the second order phase transition; in the region in-between the lines there are both the first and the second order phase transitions at different temperatures; in the upper region (orange) only the first order phase transition occurs. }     
\end{figure}

In figure~\ref{m1lowq} we show how the order parameter depends on temperature for two values of $q$ and different~$r$. The plots are shown for $\lambda=3.5$. We observe the same type of behaviour as was already reported in the Ising case ($q=2$) \cite{Sarkanych19}.
For a small number of invisible states $r<r_{c1}$, both order parameters depend on temperature continuously, leading to the second order phase transition. 
On the contrary, for large $r>r_{c2}$, only the first order transition is observed, when the magnetisation drops to zero at a particular temperature. 
Finally, for the intermediate values $r_{c1}<r<r_{c2}$, there are two transitions: first order transition between two (partially) ordered states and a second order transition when a magnetisation, that remained after the first order transition, vanishes.
This leads to the existance of two marginal values $r_{c1}(q,\lambda)$ and $r_{c2}(q,\lambda)$ that separate the regions with different criticality.  
The phase diagrams of the $q=1.5$ and $q=1.9$ Potts model with invisible states are shown in figure~\ref{fig3}. The structure of the phase diagram is similar to the Ising case ($q=2$), but the values of marginal dimensions $r_{c1}(\lambda)$ and $r_{c2}(\lambda)$ are larger. This reflects the fact that as $q$ decreases the system moves closer to the percolation case $q=1$, which is known as a very robust second order phase transition. Therefore, a larger number of invisible states $r$ is needed to change the phase transition to the first order.

\subsection{Region $q>2$}\label{III.3}

\begin{figure}[h!]
	\center
    \includegraphics[width=0.5\columnwidth]{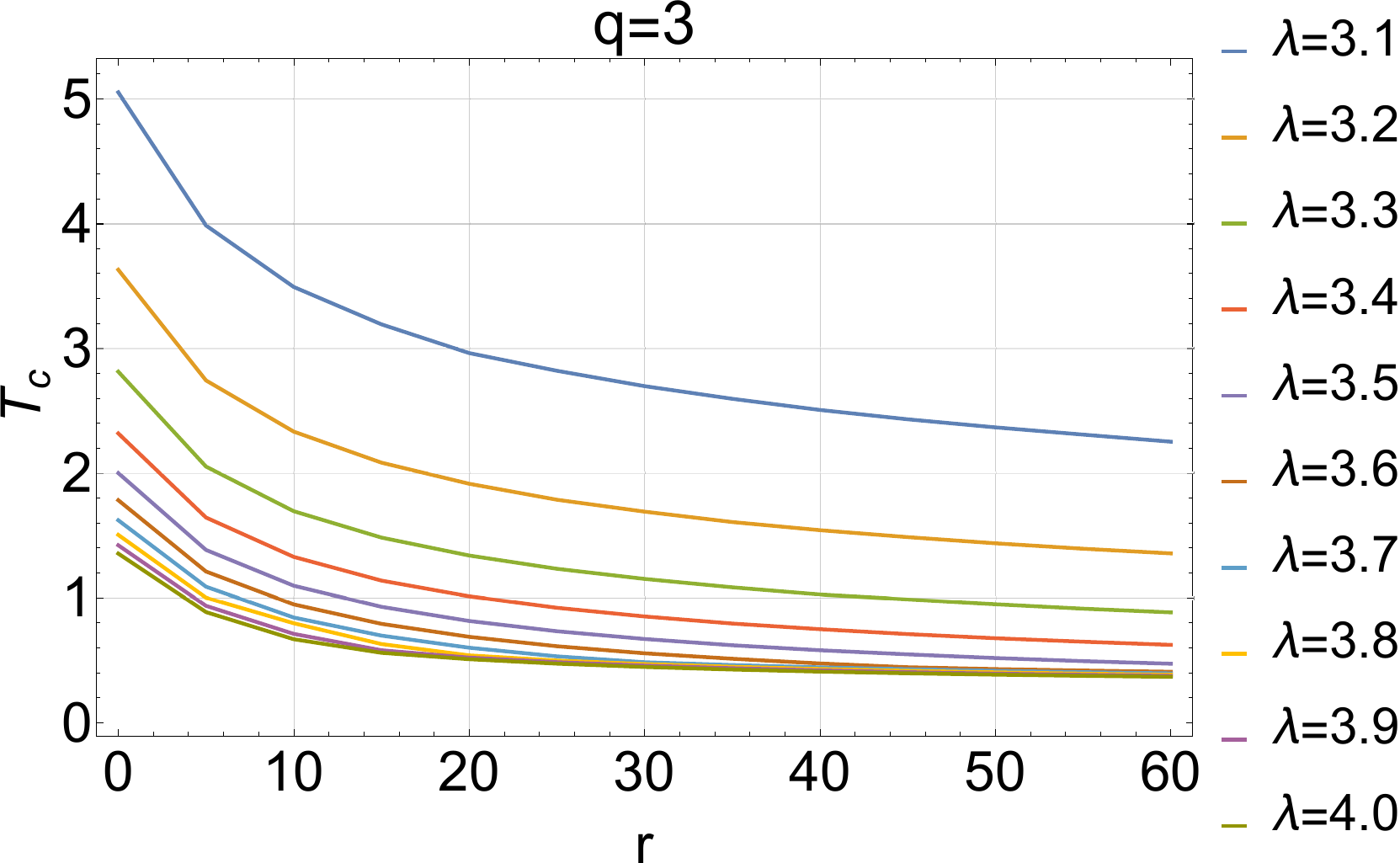}
    \caption{\label{tchighq}(Colour online) Critical temperature as a function of the number of invisible states $r$ for different values of $\lambda$. Similarly to the case $q\leqslant2$, critical temperature decreases with an increase in both $r$ and $\lambda$.}
\end{figure}

In this region, even without the presence of invisible states, the critical behaviour of the system is characterized by the marginal value $\lambda_c (q)$. 
Above this value, only the first order phase transition occurs for the standard Potts model on a scale-free network, while for $\lambda<\lambda_c(q)$ there is only a second order phase transition \cite{Krasnytska13}.
As in the previous subsection, we start by investigating critical temperature. 
In figure~\ref{tchighq} we present critical temperature as a function of the number of invisible states $r$ for different values of $\lambda$. 
The same as in the region $q\leqslant 2$, here for $q>2$ the critical temperature decreases with an increase in both $r$ and $\lambda$.

\begin{figure}[h!]
	\center
    \includegraphics[width=0.45\columnwidth]{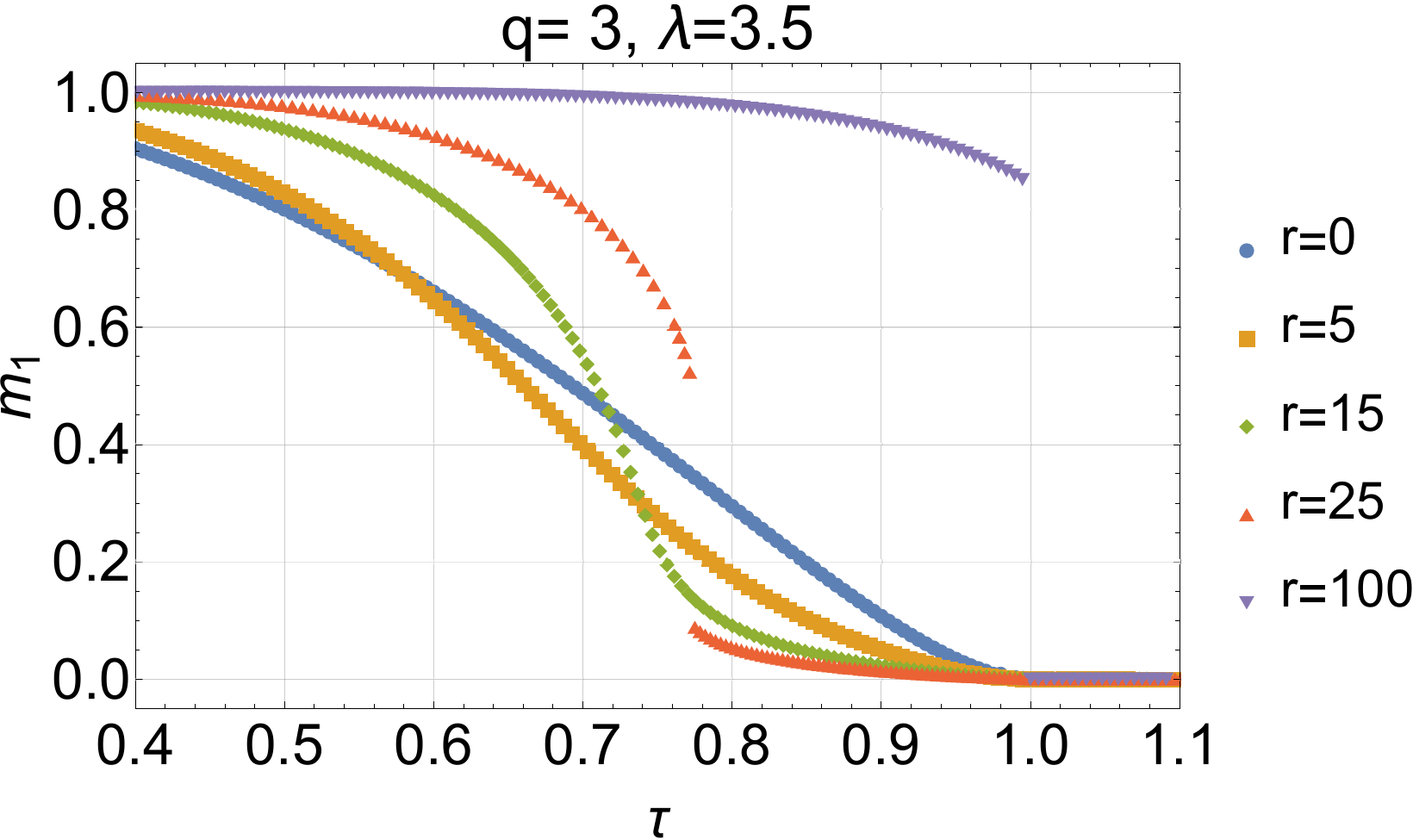} \hspace{0.5cm}
    \includegraphics[width=0.45\columnwidth]{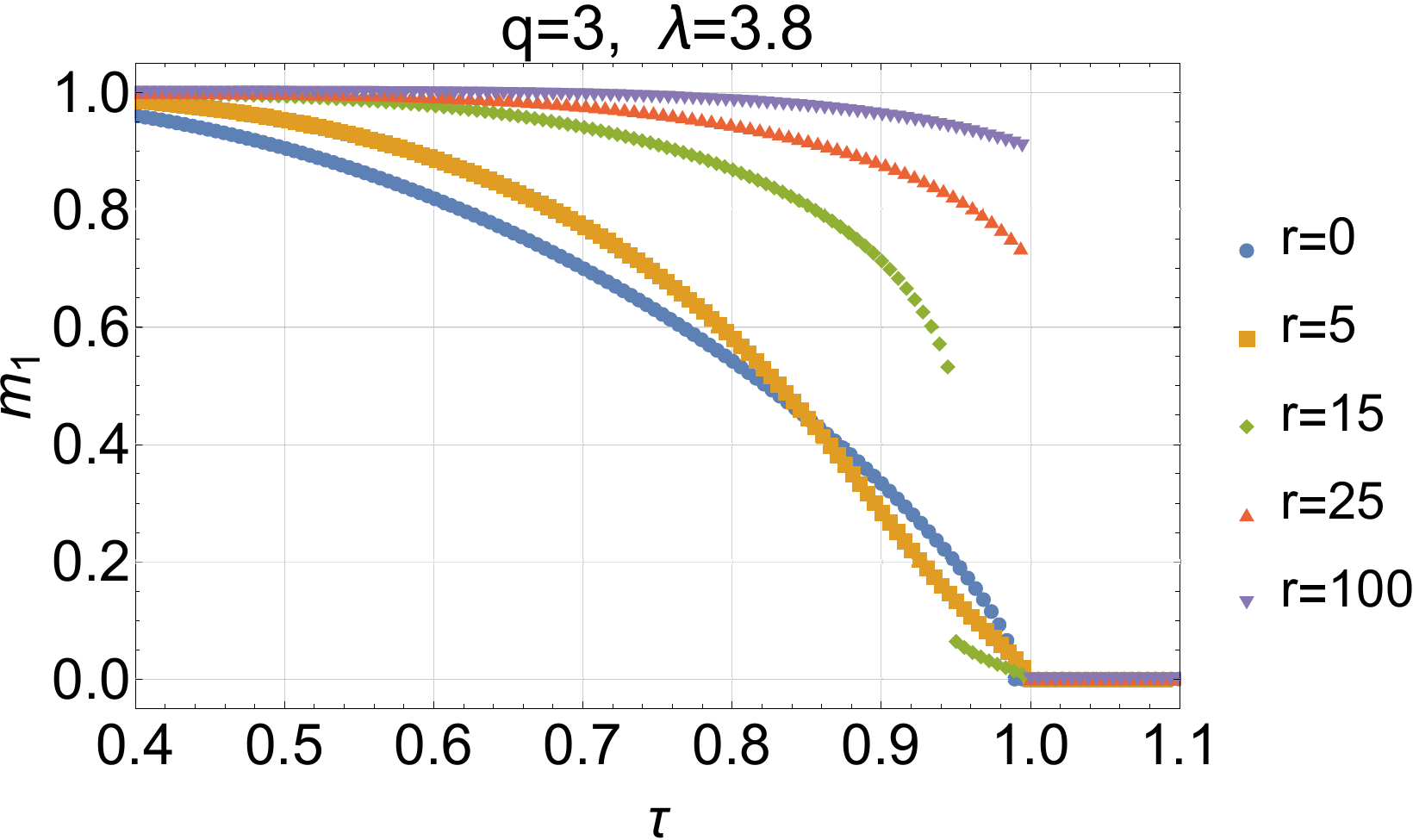}
    \caption{\label{fig4}(Colour online) First order parameter dependency on reduced temperature $\tau=T/T_c$. Left-hand plot illustrates the case $\lambda<\lambda_c(q)$ with $q=3$, $\lambda=3.5$, while on the right-hand plot $\lambda>\lambda_c(q)$ with $q=3$, $\lambda=3.8$. We see that in both of these cases gradual increase in the number of invisible states triggers a first order phase transition between two partially ordered states, while the original phase transition remains at $T_c$. When the number of invisible states is high enough, only one first order transition remains.}    
\end{figure}

Since there are two different regions depending on the value of $\lambda$, we examine how the addition of invisible states will affect the critical behaviour in both of these cases. In figure~\ref{fig4} we show the  order parameter $m_1$ dependency on the reduced temperature $\tau=T/T_c$ for $q=3$, $r=0,5,15,25,100$ and two values of $\lambda$ presenting the regions with different critical behaviour: $\lambda=3.5<\lambda_c(q=3)$ and $\lambda=3.8>\lambda_c(q=3)$.
Similarly to the case $q\leqslant2$ described in the previous subsection, we see that the small number of invisible states does not change the order of the phase transition while a large number of invisible states leads to the existence of only the first order phase transition. 
However, in the intermediate region, the situation is different. It was shown that an increase of invisible states $r$ in the  spin system with the second order phase transition triggers the first order phase transition at a lower temperature. 
In the previous papers, it was assumed that invisible states do not affect the existing first order phase transition~\cite{Krasnytska16,Tamura10}.

\begin{figure}[h!]
	\center
    \includegraphics[width=0.45\columnwidth]{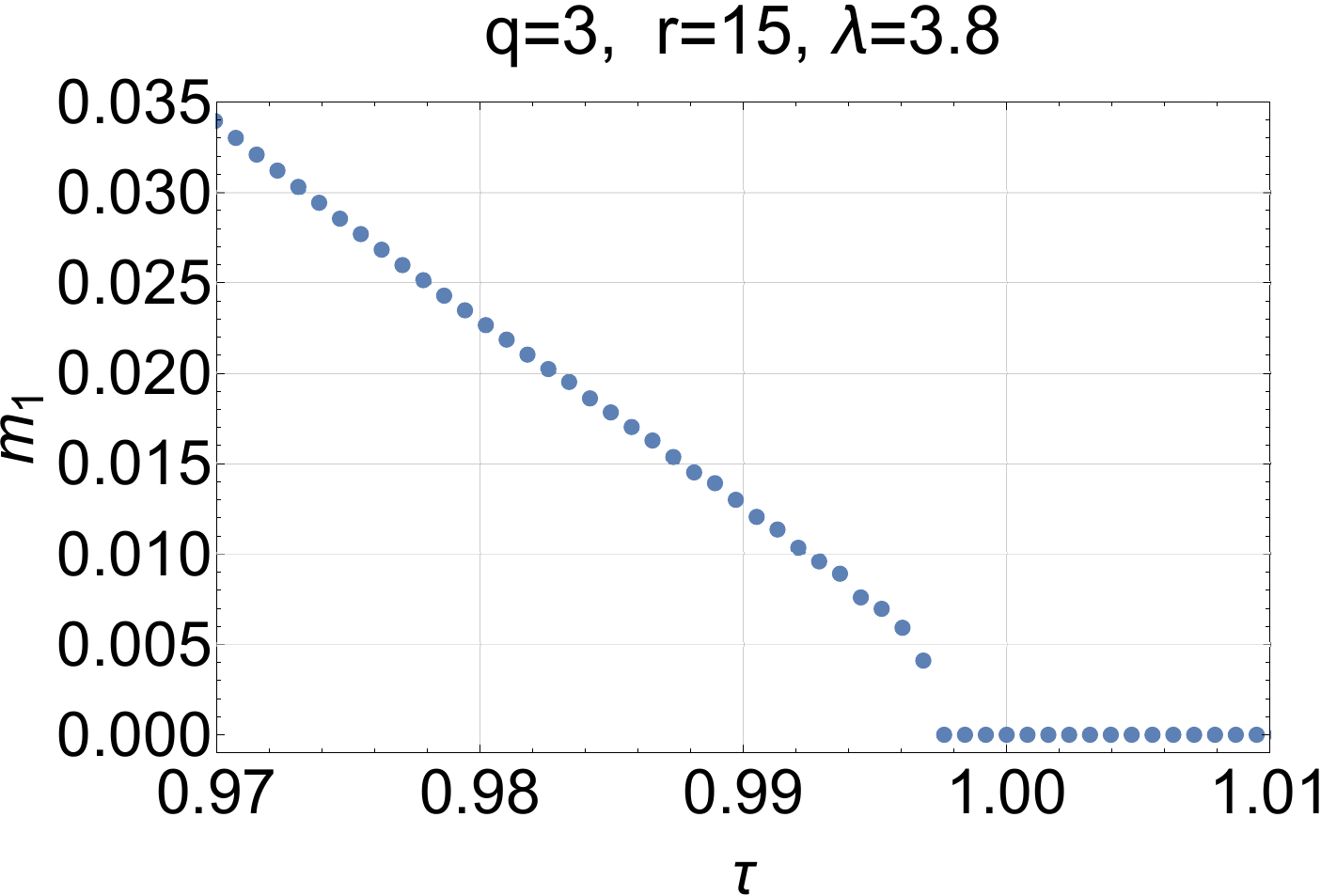} \hspace{0.5cm}
    \includegraphics[width=0.45\columnwidth]{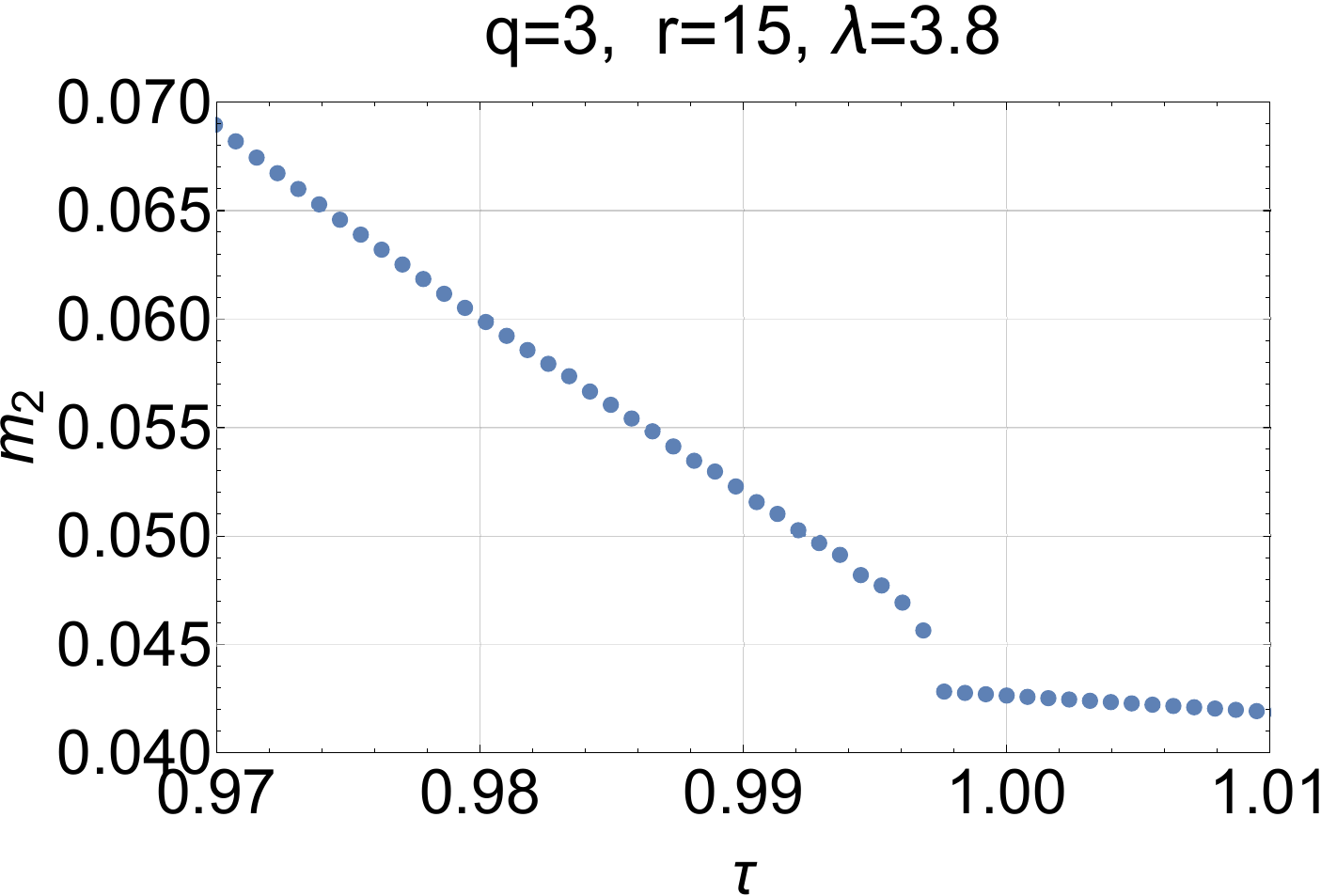}
    \caption{\label{fig4zoom}(Colour online) First and second order parameters dependency on the reduced temperature $\tau=T/T_c$ near $\tau=1$ for $q=3, r=15$ and $\lambda=3.8$. Both order parameters have jumps near the critical temperature signalling the existence of the first order phase transition.}    
\end{figure}

\begin{figure}[h]
	\center
    \includegraphics[width=0.5\columnwidth]{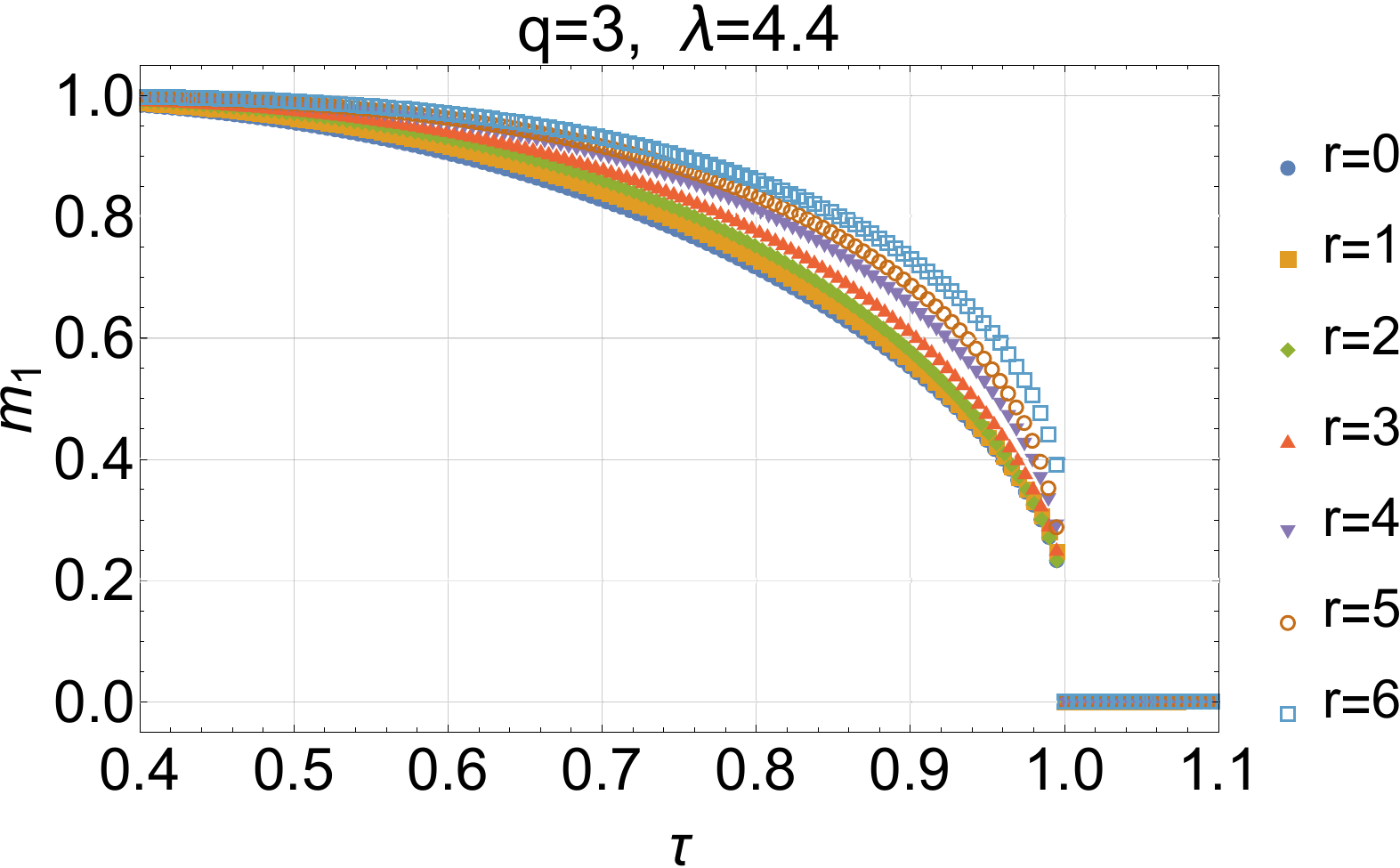}
    \caption{\label{m1q3l44}(Colour online) First order parameter dependency on the reduced temperature $\tau=T/T_c$ for $q=3$, $\lambda=4.4$ and $r$ ranging from 0 to 6. The increase in the number of invisible states does not trigger an additional first order phase transition. However, it makes the transition sharper.}    
\end{figure}

Here, we see that in the first order regime $\lambda>\lambda_c(q)$ we also obtain an additional first order transition at a lower temperature.
From figure~\ref{fig4} it might be unclear what is happening with the initial first order phase transition when invisible states are added. 
To illustrate this, in figure~\ref{fig4zoom} we show the dependency of both order parameters on temperature near the critical point $\tau=1$. Both order parameters experience jumps signalling the existence of the first order phase transition. 
Therefore, the system is characterised by two consecutive first order transitions. This type of critical behaviour had not been reported before for the Potts model with invisible states. This phenomenon is observed only in the limited area of $\lambda$. Apparently, there is some $\lambda_u$ that limits the region where two consecutive first order phase transitions occur. 
In figure~\ref{m1q3l44} we demonstrate the first order parameter dependency on the reduced temperature for $q=3$, $\lambda=4.4$ and~$r$ ranging from 1 to 6. We see that with $r$ increasing, the existing first order phase transition only gets sharper and no additional transition occurs. For $q=3$, we have found that $4.1<\lambda_u<4.2$.

\begin{figure}[h!] 
	\center
    \includegraphics[width=0.6\columnwidth]{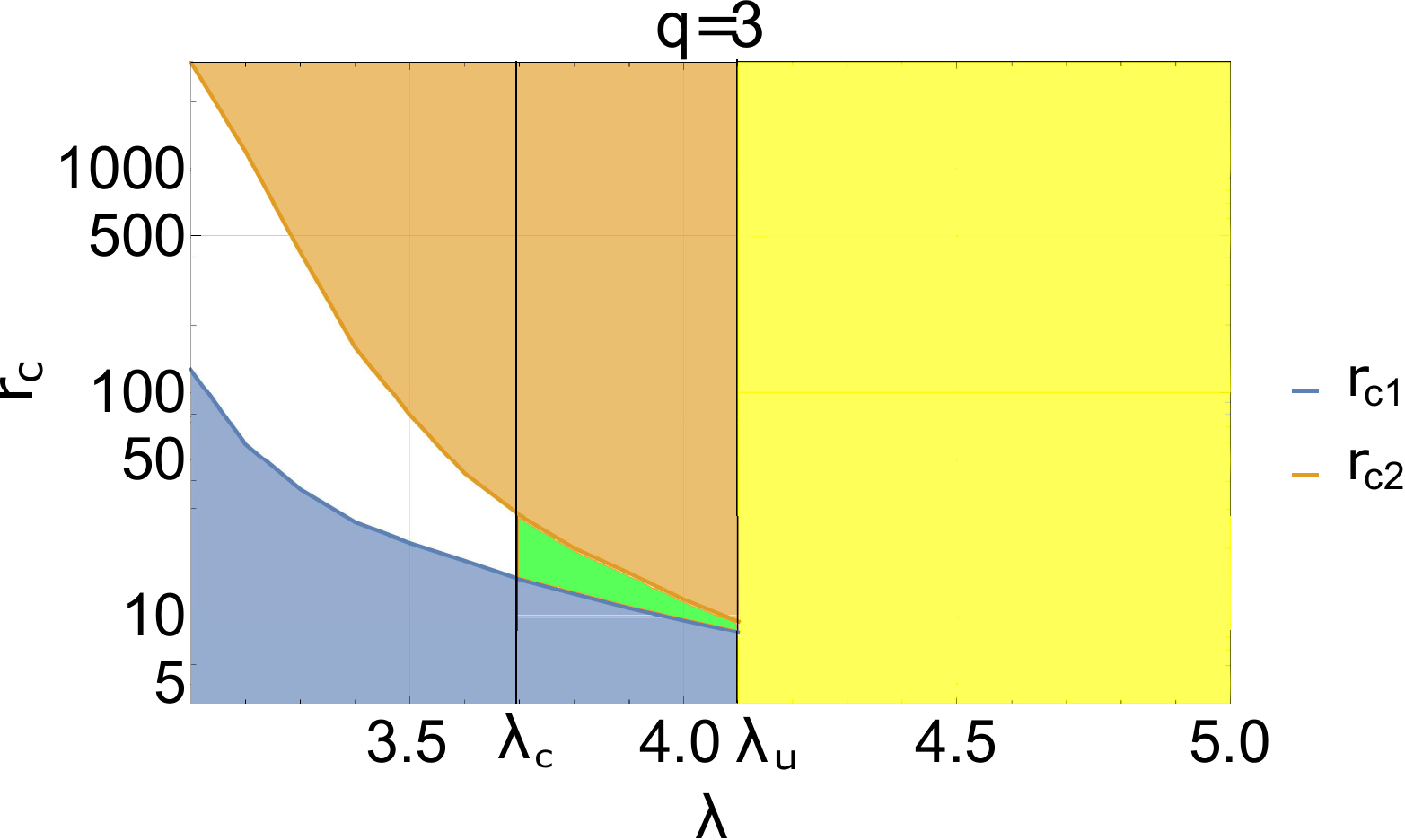} 
    \caption{\label{fig5}(Colour online) Phase diagram of the Potts model with invisible states on a scale free network for $q=3$. 
    For $\lambda<\lambda_c(q)$, the phase diagram is qualitatively the same as in the $q\leqslant2$ case. At a large number of invisible states $r>r_{c2}$ (dark yellow region), the first order phase transition occurs. If the amount of invisible states is small, the system is described by the second order phase transition behaviour  (the blue region), and in between --- white region --- both the first and the second order phase transitions occur
at different temperatures. However, it is different for $\lambda>\lambda_c(q)$:  two first order phase transitions
occur (green and dark yellow regions). In the region $\lambda>\lambda_u$, any amount of invisible states is incapable of affecting the critical behaviour, the single first order transition occurs (light yellow).}
     
\end{figure}

Summarizing the results on the phase diagram (see figure~\ref{fig5}), we may conclude that even in the case $q>2$ the system is still characterised by two marginal dimensions $r_{c1}(\lambda)$ and $r_{c2}(\lambda)$, although the critical behaviour in the regions into which these marginal dimensions split the phase diagram is $\lambda$-dependent. For $\lambda<\lambda_c(q)$ adding invisible states $r$ has the same effect as in the previous case, i.e., for small values $r<r_{c1}(\lambda)$ there is only the second order phase transition; for $r_{c1}(\lambda)\leqslant r \leqslant r_{c2}(\lambda)$, there are both the first and the second order phase transitions occurring at different temperatures; for $r>r_{c2}(\lambda)$, only the first order transition remains. For $\lambda_c(q)<\lambda<\lambda_u(q)$, the behaviour is very similar with a sole difference that the second order phase transition is replaced by the first order transition, but two marginal values $r_{c1}(\lambda)$ and $r_{c2}(\lambda)$ remain. Finally, for   $\lambda_u(q)<\lambda$, there is only the first order phase transition with the jump in the order parameter dependent on the number of invisible states.  The existence of the region where two first order phase transitions occur suggests that the topological and entropic effects on the phase transitions are to a certain extent independent of each other. 
Invisible states induce the corresponding first order transition regardless of the nature of the transition  observed at the critical point where the first order parameter vanishes $m_1=0$. 
On the other hand, when a large number of invisible states is added to the model, the transition at the critical point is always a first order transition for all values of $\lambda$ where transition can occur.

\section{Conclusions}
\label{IV} 
 
In this paper, we sum up the analysis of the critical behaviour of the Potts model with invisible states on an annealed scale-free network. Using the inhomogeneous mean-field approximation, we confirmed our previously obtained results and conclusions \cite{Sarkanych18,Sarkanych19,Krasnytska13}, namely that the number of invisible states can change the universality class of the standard models on a complete graph or even on a scale-free network \cite{Sarkanych19} when the degree distribution
decay exponent plays a similar role. 
After numerical analysis of the free energy of the Potts model with invisible states on a scale-free network, we can conclude that $q$, $r$ and $\lambda$ play a role of global parameters that influence the critical behaviour of the system. We investigated the interplay of the parameters in different regions of $q$, $r$, $\lambda$ and sum it up on the phase diagrams. 
Depending on $q$ and $\lambda$, the critical behaviour changes.
Unlike in reference~\cite{Sarkanych17} we cannot analyze the behaviour of the system for a negative number of invisible states. This is due to the method we applied. Negative $r$ could lead to complex values of the free energy (\ref{ff0}) that cannot be minimized.

For strongly correlated network region  $2<\lambda \leqslant 3$ it was shown in subsection \ref{III.1} that any large number of invisible states is incapable of breaking the order, the system stays ordered at any temperature.  In section \ref{III.2} for $1<q\leqslant 2$, we obtained a phase diagram similar to the one for the Ising case (see figure~\ref{fig3}).
The phase diagram in the $(r,\lambda)$ plane is divided by two marginal values $r_{c1}(\lambda)$ and $r_{c2}(\lambda)$ into three regions with different critical behaviour.  
In the lower region $r<r_{c1}(\lambda)$ the system undergoes only the second order phase transition; in the region in-between the lines there are both first and second order phase transitions at different temperatures; for a large number of invisible states $r>r_{c2}(\lambda)$, only the first order phase transition occurs. The last region of $q>2$ (see subsection \ref{III.3}) is the most interesting.  In this region, two marginal values of $\lambda_c$  and $\lambda_u$ appear. 
For $\lambda<\lambda_c$, the phase diagram (figure~\ref{fig5}) is similar to what is observed for the $q\leqslant 2$ case in figure~\ref{fig3}.
In the region $\lambda_c<\lambda<\lambda_u$, the phase diagram is also characterized by two marginal values $r_{c1}(\lambda)$ and $r_{c2}(\lambda)$. For $r<r_{c1}(\lambda)$, the system is characterized by the first order phase transitions same as in $r=0$ case. 
In the region $r_{c1}(\lambda)<r<r_{c2}(\lambda)$, there are two first order phase transitions at different temperatures. At a lower temperature, the transition is between two partially ordered states, while at a higher temperature, the system becomes fully disordered $m_1=0$. 
Lastly, for $r>r_{c2}(\lambda)$, there is only a single first order transition between an ordered and disordered state.
In the region $\lambda>\lambda_u$ any amount of invisible states is incapable of affecting the critical behaviour. There is only a single first order transition regardless of the value of $r$.

We observed that the addition of invisible states triggers an additional first order phase transition that, for a certain range of $r$, exists alongside the phase transition that existed at $r=0$. This implies that the topological and entropic effects on the phase transition are independent.
The presence of non-integer marginal dimensions is ubiquitous in the criticality of complex systems. It should be mentioned that a topology presented by the network structure determined by node degree distribution decay exponent $\lambda$ dominates over the entropic influence, although the effect of an additional first order phase transition should be analyzed more in detail. 

The value of $\lambda_c$ we observed in the $q>2$ case fully coincides with the value obtained in reference~\cite{Krasnytska13}. It was obtained from the condition that the coefficient of the $\lambda$-dependent term in the free energy expansion changes the sign and requires taking the next term of the expansion into consideration. In reference~\cite{Krasnytska13} it was also shown that there are higher values of $\lambda$ that can change the sign of the corresponding coefficient, but since for such  $\lambda$ the expansion term is irrelevant for the phase diagram, these values are omitted. Interestingly enough, the value $\lambda_u$ that we observed in this paper is similar to the second lowest value of~$\lambda$ that changes the sign of the leading $\lambda$-dependent coefficient in the free energy expansion over the order parameter for the standard Potts model on an annealed scale-free network with $r=0$. Whether this is just a coincidence or has a deeper meaning needs an additional investigation.

\section*{Acknowledgement}
We acknowledge Yurij Holovatch for fruitful discussions and useful
comments. Our work was supported in part by the National Academy of Sciences of Ukraine [the grant for research laboratories/groups of young scientists No 07/01-2022(4)] (M.~K.) and National Academy of Sciences of Ukraine,
project No.~654~1230 (P.~S.). 
M.~K. acknowledges support of the PAUSE programme and the hospitality of
the Laboratoire de Physique et Chimie Th\'eoriques,  Universit\'e de Lorraine (France).
 
We thank the Ukrainian Army for the possibility to perform this research work.

\ukrainianpart
 
\title{╠юфхы№ ╧юЄЄёр ч эхтшфшьшьш ёЄрэрьш эр схчьрё°Єрсэ│щ ьхЁхц│}
\author{╧. ╤рЁърэшў\refaddr{label1,label2}, ╠. ╩ЁрёэшЎ№ър\refaddr{label1,label2,label3}}
\addresses{
	\addr{label1}▓эёЄшЄєЄ Ї│чшъш ъюэфхэёютрэшї ёшёЄхь ═└═ ╙ъЁр┐эш, тєы. ╤т║эЎ│Ў№ъюую, 1, 79011 ╦№т│т, ╙ъЁр┐эр
	\addr{label2}╤яiтяЁрЎ  \textbf{L}$^4$ i ╩юыхфц фюъЄюЁрэЄiт чi ёЄрЄшёЄшўэю┐ Їiчшъш ёъырфэшї ёшёЄхь,
	╦ щяЎiуЦ╦юЄрЁшэуi Ц╦№тiтЦ╩ютхэЄЁi, ктЁюяр
	\addr{label3} ╦рсюЁрЄюЁ│  Ї│чшъш Єр ї│ь│┐, ╙э│тхЁёшЄхЄ ╦юЄрЁшэу│┐, ┬рэфєЁт-ы║-═рэё│, 54506, ╘ЁрэЎ│ }
\makeukrtitle

\begin{abstract}
\tolerance=3000%
─ы  яю ёэхээ  ьруэ│Єэшї Їрчютшї яхЁхїюф│т ўхЁхч ъюэъєЁхэЎ│■ хэхЁу│┐ Єр хэЄЁюя│┐ сєыю чряЁюяюэютрэю Ё│чэ│ ьюфхы│. ╬фэр ч эшї -- $q$-ёЄрэютр ьюфхы№ ╧юЄЄёр ч эхтшфшьшьш ёЄрэрьш.
┬ Ёрьърї Ў│║┐ ьюфхы│ ттюф Є№ $r$ эхтшфшьшї ёЄрэ│т Єръшь ўшэюь, ∙ю ёя│э,  ъшщ яхЁхсєтр║ є сєф№- ъюьє ч эшї, эх тчр║ьюф│║ ч │э°шьш ёя│эрьш є ёшёЄхь│.
╠ш Ёючуы фр║ью ьюфхы№ эр т│фярыхэ│щ схчьрё°Єрсэ│щ ьхЁхц│, фх │ьют│Ёэ│ёЄ№, ∙ю тшярфъютю юсЁрэшщ тєчюы ьр║ ёЄєя│э№ $k$ юяшёє║Є№ё  ёЄхяхэхтю-ёярфэшь чръюэюь $P(k)\propto k^{-\lambda}$.
═р°│ Ёхчєы№ЄрЄш я│фЄтхЁфцє■Є№, ∙ю $q,r$ Єр $\lambda$ т│ф│уЁр■Є№ Ёюы№ уыюсры№эшї ярЁрьхЄЁ│т, ∙ю тшчэрўр■Є№ ъЁшЄшўэє яютхф│эъє ёшёЄхьш.
╟рыхцэю т│ф ┐ї чэрўхэ№, Їрчютр ф│руЁрьр ёъырфр║Є№ё  │ч ЄЁ№юї юсырёЄхщ ч Ё│чэю■ ъЁшЄшўэю■ яютхф│эъю■.
╬фэръ, ёы│ф чрєтрцшЄш, ∙ю тяышт Єюяюыюу│┐, яЁхфёЄртыхэшщ уЁрэшўэшьш чэрўхээ ьш $\lambda_c(q)$, фюь│эє║ эрф хэЄЁюя│щэшь ЇръЄюЁюь, ∙ю тшчэрўр║Є№ё  ъ│ы№ъ│ёЄ■ эхтшфшьшї ёЄрэ│т $r$.

 
\end{abstract}

\lastpage

\begin{thebibliography}{99}

\bibitem{Ising}
 Ising E., Z. Phys., 1925, {\bf 31}, 253, \doi{10.1007/BF02980577}.
\bibitem{Onsager}
 Onsager L., Phys. Rev., 1944, {\bf 65}, 117, \doi{10.1103/PhysRev.65.117}.
\bibitem{El-Showk}
 El-Showk S.,  Paulos M.~F., Poland D., Rychkov S., Simmons-Duffin D., Vichi A.,
{Phys. Rev. D}, 2012, {\bf 86}, No.~2, 025022, \doi{10.1103/PhysRevD.86.025022}.
 \bibitem{networks_1}
 Albert  R.,  Barab\'asi A.~L., {Rev. Mod. Phys.}, 2002, {\bf 74}, 47, \doi{10.1103/RevModPhys.74.47}.
\bibitem{networks_2}
 Dorogovtsev S.~N., Mendes J.~F.~F., Evolution of Networks: From
    Biological Networks to the Internet and WWW, { Oxford University
        Press}, Oxford, 2003.
\bibitem{networks_3}
    Holovatch  Yu., von~Ferber C., Olemskoi  A.,  Holovatch T., Mryglod O., Olemskoi I.,  Palchykov V., {J.~Phys.~Stud.}, 2006, {\bf 10}, 247, \doi{10.30970/jps.10.247}, (in Ukrainian). 
    \bibitem{networks_4}
     Barrat A., Barthelemy  M.,  Vespignani A., Dynamical Processes on Complex Networks, {Cambridge University Press}, 2008.
    \bibitem{networks_5}
     Newman M., Networks: An Introduction, {Oxford University
        Press}, 2010.
    \bibitem{Dorogovtsev08}
    Dorogovtsev S.~N., Goltsev A. V., {Rev. Mod. Phys.}, 2008, {\bf 80}, 1275, \doi{10.1103/RevModPhys.80.1275}.
	
\bibitem{Leone02}
     Leone M.,  V\'azquez A., Vespignani A.,  Zecchina R., Eur. Phys. J. B, 2002, {\bf 28}, 191, \doi{10.1140/epjb/e2002-00220-0}.
\bibitem{Igloi02}
    Igloi F., Turban L., {Phys. Rev. E}, 2002, {\bf 66}, 036140, \doi{10.1103/PhysRevE.66.036140}.

    \bibitem{Dorogovtsev02}
     Dorogovtsev S.~N., Goltsev A.~V., Mendes J.~F.~F., {Phys. Rev. E}, 2002, {\bf 66},  016104,\\ \doi{10.1103/PhysRevE.66.016104}.
    \bibitem{Tamura10}
     Tamura R., Tanaka S., Kawashima N., {Prog. Theor. Phys.}, 2010, {\bf 124}, No.~2, 381, \doi{10.1143/PTP.124.381}.
    \bibitem{Tamura11}
    Tanaka S., Tamura R., Kawashima N., {J. Phys. Conf. Ser.}, 2011, {\bf 297}, 012022, \doi{10.1088/1742-6596/297/1/012022}.
    \bibitem{Tanaka11a}
     Tanaka S., Tamura R., {J. Phys. Conf. Ser.}, 2011,  {\bf 320}, 012025, \doi{10.1088/1742-6596/320/1/012025}.
	\bibitem{Johnston13}
     Johnston D.~A., Ranasinghe R.~P.~K.~C.~M., {J. Phys. A: Math. Theor.}, 2013, {\bf 46}, 225001, \doi{10.1088/1751-8113/46/22/225001}.
    \bibitem{Mori12}
     Mori T., {J. Stat. Phys.}, 2012, {\bf 147}, 1020, \doi{10.1007/s10955-012-0511-0}.
    \bibitem{Enter11a}
    van Enter A.~C.~D., Iacobelli G., Taati S., {Prog. Theor. Phys.}, 2011, {\bf 126},  983, \doi{10.1143/PTP.126.983}.
    \bibitem{Enter11b}
    van Enter A.~C.~D., Iacobelli G., Taati S., {Rev. Math. Phys.}, 2012, {\bf 24}, No.~02, 1250004,\\ \doi{10.1142/S0129055X12500043}.
    \bibitem{Ananikian13}
     Ananikian N., Izmailyan N. Sh., Johnston D. A., Kenna R.,  Ranasinghe~R.~P.~K.~C.~M.,
    {J. Phys. A: Math. Theor.}, 2013, {\bf 46}, 385002, \doi{10.1088/1751-8113/46/38/385002}.
    \bibitem{Sarkanych17}
    Sarkanych P., Holovatch Yu., Kenna R., {Phys. Lett. A}, 2017, {\bf381}, No.~41, 3589--3593,\\ \doi{10.1016/j.physleta.2017.08.063}.
     \bibitem{Rojas20}
    Rojas O., {Braz. J. Phys.}, 2020, {\bf 50}, 675--686, \doi{10.1007/s13538-020-00773-8}.
     \bibitem{Ninio76}
 Ninio F., {J. Phys. A: Math Gen.}, 1976, {\bf 9}, 1281, \doi{10.1088/0305-4470/9/8/017}.
    
 \bibitem{Kiraly19}
 Kir\'aly B., Ph.D. Thesis, Budapest University of Technology and Economics, Hungarian Academy of Science, Budapest, 2019.

  \bibitem{Korbel21}
      Korbel J., Lindner S.~D., Hanel R., Thurner S.,  {Nat. Commun.}, 2021, {\bf 12}, 1127, \doi{10.1038/s41467-021-21272-7}.
\bibitem{Schreiber22}
 Schreiber N., Cohen R., Amir G., Haber S., J. Stat. Mech.: Theory Exp., 2022, {\bf 2022}, 043205, \doi{10.1088/1742-5468/ac603a}.
    \bibitem{Krasnytska16}
      Krasnytska M., Sarkanych P., Berche B., Holovatch Yu., Kenna R., {J. Phys. A: Math. Theor.}, 2016, {\bf 49}, 255001,\\ \doi{10.1088/1751-8113/49/25/255001}.
    
 \bibitem{Sarkanych19}
   Sarkanych P., Krasnytska M.,  {Phys. Lett. A}, 2019, {\bf 383}, 125844, \doi{10.1016/j.physleta.2019.125844}.
    
    \bibitem{Lee09}
 Lee S.~H., Ha M., Jeong H., Noh J.~D., Park H., {Phys.
Rev. E}, 2009, {\bf 80}, 051127,\\ \doi{10.1103/PhysRevE.80.051127}.
 \bibitem{Krasnytska13}
     Krasnytska M., Berche B., Holovatch Yu., {Condens. Matter Phys.}, 2013, {\bf 16},  23602, \doi{10.5488/CMP.16.23602}.
    \bibitem{Aiello2000}
         Aiello W.,  Chung F.,  Lu L.,  {Exp. Math.}, 2001, {\bf 10}, 53, \doi{10.1080/10586458.2001.10504428}.
    
   \bibitem{SimulatedAnnealing}
    Dueck G., Scheuer T., 
    J. Comput. Phys., 1990, {\bf 90}, No.~1, 161--175, \doi{10.1016/0021-9991(90)90201-B}.
    
    \bibitem{Sarkanych18}
    Sarkanych P., Holovatch Yu., Kenna R., {J. Phys. A: Math. Theor.}, 2018, {\bf 51},  505001, \doi{10.1088/1751-8121/aaea02}.
      
\end{thebibliography}
\end{document}